# Comment on
"Chebyshev finite difference method for the effects of variable viscosity and variable thermal conductivity on heat transfer to a micro-polar fluid from a non-isothemal stretching sheet with suction and blowing", by S.N. Odda and A.M. Farhan [Chaos, Solitons & Fractals, 2006, vol. 30, pp. 851-858]

Asterios Pantokratoras
Associate Professor of Fluid Mechanics

School of Engineering, Democritus University of Thrace,
67100 Xanthi – Greece
e-mail:apantokr@civil.duth.gr

In the above paper the authors treat the boundary layer flow of a micropolar fluid along a horizontal flat plate with blowing or suction. The fluid viscosity and thermal conductivity are assumed functions of temperature. The boundary layer equations are transformed into ordinary ones and subsequently are solved using the Chebyshev finite difference method. However, there are some deficiencies and errors in this paper which are presented below:

1. In the introduction it is mentioned that the theory of micro-polar fluids can be used to explain the flow of colloidals, liquid crystals, animal blood etc. However the rest of the paper is referred to air, water and lubrication oils and results are presented only for Prandtl number 0.7 (air). These three fluids are not micro-polar and there is no relationship between them and colloidals, liquid crystals and animal blood. The authors used a theory that is not valid for the fluids used (air, water and oils).

2. In the section governing equations it is mentioned twice that the fluid thermal conductivity is assumed to be an inverse linear function of temperature whereas the equation used in the paper is a linear function of temperature.

3. In the transformed energy equation (9) the Prandtl number appears in two terms and has been assumed constant across the boundary layer. All the presented results concern Pr=0.70.

However, the Prandtl number is a function of viscosity and thermal conductivity and these quantities are functions of temperature. Taking into account that temperature varies across the boundary layer, the Prandtl number varies, too. The assumption of constant Prandtl number inside the boundary layer leads to unrealistic results (Pantokratoras, 2004, 2005). The problem can be treated properly either considering the Prandtl number as a variable in the transformed equations (Saikrishnan and Roy, 2003) or with the direct solution of the initial boundary layer equations and treating the fluid properties as functions of temperature (Pantokratoras, 2004, 2005).

4. In the section governing equations it is mentioned that the mass transfer parameter is positive for injection and negative for suction. It is well known in fluid mechanics that injection thickens the boundary layer and suction reduces the boundary layer thickness ( White, 1991, page 251, Schlichting and Gersten, 2003, pages 297,299). However the opposite happens in figures 5 and 7 of Odda and Farhan (2006). The positive mass transfer parameter (injection) corresponds to thin profiles and negative mass transfer parameter (suction) corresponds to thick profiles.

5. It is known in boundary layer theory that velocity and temperature profiles approach the ambient fluid conditions asymptotically and do not intersect the line which represents the boundary conditions as happens in some profiles of figures 2 and 9 (See for example the velocity profile for a moving plate on page 177 by Schlichting and Gersten, 2003). Especially the temperature profile in figure 9 which corresponds to $\gamma=-1$ is almost a straight line. It is clear that these profiles, which do not approach the horizontal axis asymptotically and intersect it, are truncated due to a small calculation domain used. The authors used for all cases a calculation domain with $\eta_{max}=10$. This calculation domain was sufficient to capture the majority of profiles but insufficient to capture some temperature profiles. This problem can be solved by using a wider calculation domain, greater than 10.

Taking into account the above arguments it is clear that the results included in the paper by Odda and Farhan (2006) are wrong both from a theoretical and practical point of view.